\documentclass[11pt]{amsart}
\usepackage[margin=1in]{geometry}                % See geometry.pdf to learn the layout options. There are lots.
\usepackage{graphicx}
\usepackage{enumitem}
\usepackage{amssymb}
\usepackage{color}
\usepackage{wrapfig}
\usepackage{appendix}
\usepackage[rightcaption]{sidecap}
\usepackage{caption}
\usepackage[colorlinks={true},pdftex]{hyperref}
\hypersetup{citecolor={blue}, filecolor={blue}, linkcolor={blue}, urlcolor={blue}}
\graphicspath{{Figures/}{../Figures/}}
\newcommand{\hname}[1]{#1}

\usepackage[numbers,sort&compress]{natbib}

\newcommand{\iso}[2]{${}^{#1}\mathrm{#2}$}
\newcommand{\ut}[1]{\,\mathrm{#1}}

\newcommand{\xe}{$^{129}$Xe\ }
\newcommand{\mf}[1]{\mathrm{#1}}
\newcommand{\E}[1]{$\times 10^{#1}$}

\begin{document}
%\maketitle

\begin{center}
\huge{Comagnetometer probes of dark matter and new physics}
\end{center}
%\vspace{0.03in}
\begin{center}
\large{W. A. Terrano and M. V. Romalis  \\ Physics Department, Princeton University}
\end{center}

\begin{center}

Modern comagnetometry is -- in absolute energy units -- the most sensitive experimental technique for measuring the 
energy splitting between quantum states, with certain implementations measuring the nuclear spin-up/spin-down splitting at the $\mathrm{10^{-26}\,eV}$ level.  
By measuring and 
subtracting the leading magnetic effects on the spins, comagnetometry can be used to study {non-standard-model} spin interactions.  New physics 
scenarios that comagnetometers can probe include EDMs, violations of Lorentz invariance, Goldstone 
bosons from new high-energy symmetries, spin-dependent and CP-violating long-range forces, and axionic dark matter. We 
describe the many implementations that have been developed and optimized for these applications, and 
 consider the prospects for improvements in the technique.  Based purely on existing 
technology, there is room for several orders of magnitude in further improvement in statistical sensitivity.  
We also evaluate sources of systematic error and instability that may limit attainable improvements.
\end{center}

\vspace{0.03in}

\section{Introduction and history of the field}
Following the discovery of nuclear spins in 1933\cite{Estermann1933} and the demonstration of control over them via their magnetic 
moments in 1938\cite{Rabi1938}  -- accomplishments earning the 1943 Nobel prize for Otto Stern and the 1944 Nobel prize for 
Isidor Rabi respectively -- nuclear magnetic resonance burgeoned as a technique in condensed matter and medical physics, beginning with 
the experiments of Bloch and Purcell\cite{Bloch1946, Purcell1946} in 1946 for which they shared the 1952 Nobel prize.  The first use of nuclear spins for fundamental physics\footnote{Ramsey used bare neutrons for a neutron EDM measurement in 1951 that went unpublished until 1957, when the discovery of parity violation convinced them to publish their negative result.} -- and  with it the 
development 
of comagnetometry -- took place in 1960 when Hughes\cite{Hughes1960} and Drever\cite{Drever1961} 
compared the magnetic resonances of $^7\mathrm{Li}$ and a proton at different orientations of 
the 
$^7\mathrm{Li}$ quadrupole relative to the galactic center.  

A challenge when searching for new physics with spin-dependent couplings is that the magnetic 
interactions of the spins dwarf the anticipated new physics signals by several orders of magnitude.  The 
simplest {nuclear-spin} Hamiltonian for 
a system with total angular momentum $\vec{F}$ is
\begin{equation}
\mathcal{H}_\mathrm{spin} = \mathcal{H}_\mathrm{mag} + \mathcal{H}_\mathrm{BSM} + ... = \vec{\mu}_N \cdot \vec{B} + \vec{\sigma}_N \cdot \vec{\beta} + ...
\end{equation}
with the additional terms depending on the specific comagnetometer implementation.  Magnetic 
interactions  
$\mathcal{H}_\mathrm{mag}$ are described by the magnetic moment of the 
nucleus $\vec{\mu}_N = \mu \vec{F}/F$ and the magnetic field $\vec{B}$.  Beyond-the-Standard-Model 
interactions  $\mathcal{H}_\mathrm{BSM}$ are described by the spin moment of the nucleus
$\hat{\sigma}_N = \vec{F}/F$ and an effective 
field $\vec{\beta}$ appropriate to the new coupling of interest. For instance, in an electric-dipole-moment 
search $\vec{\beta} = d_\mathrm{N}\vec{E} $ with $\vec{E}$ the applied electric field,  and $d_\mathrm{N}$ 
the searched for EDM.  

To extract $\mathcal{H}_\mathrm{BSM}$  from  $\mathcal{H}_\mathrm{spin}$ the experimentalist typically 
applies a characteristic time-dependence to $\vec{\beta}(t)$.  Even so, the effect of fluctuations in $\mathcal{H}_\mathrm{mag}$  must be 
suppressed to eliminate correlations -- coincidental or systematic -- between the time-dependences of $\vec{\beta}$ and $\vec{B}$.  Comagnetometry 
achieves this by comparing spins with differing $\vec{\mu}$ or differing $\vec{\beta}$.  Other experimental approaches for studying similar types of new physics 
have been used as well, as summarized in a recent review,  for example\cite{Safronova2018}.

\begin{figure}[htp]
\scalebox{0.35}{\includegraphics{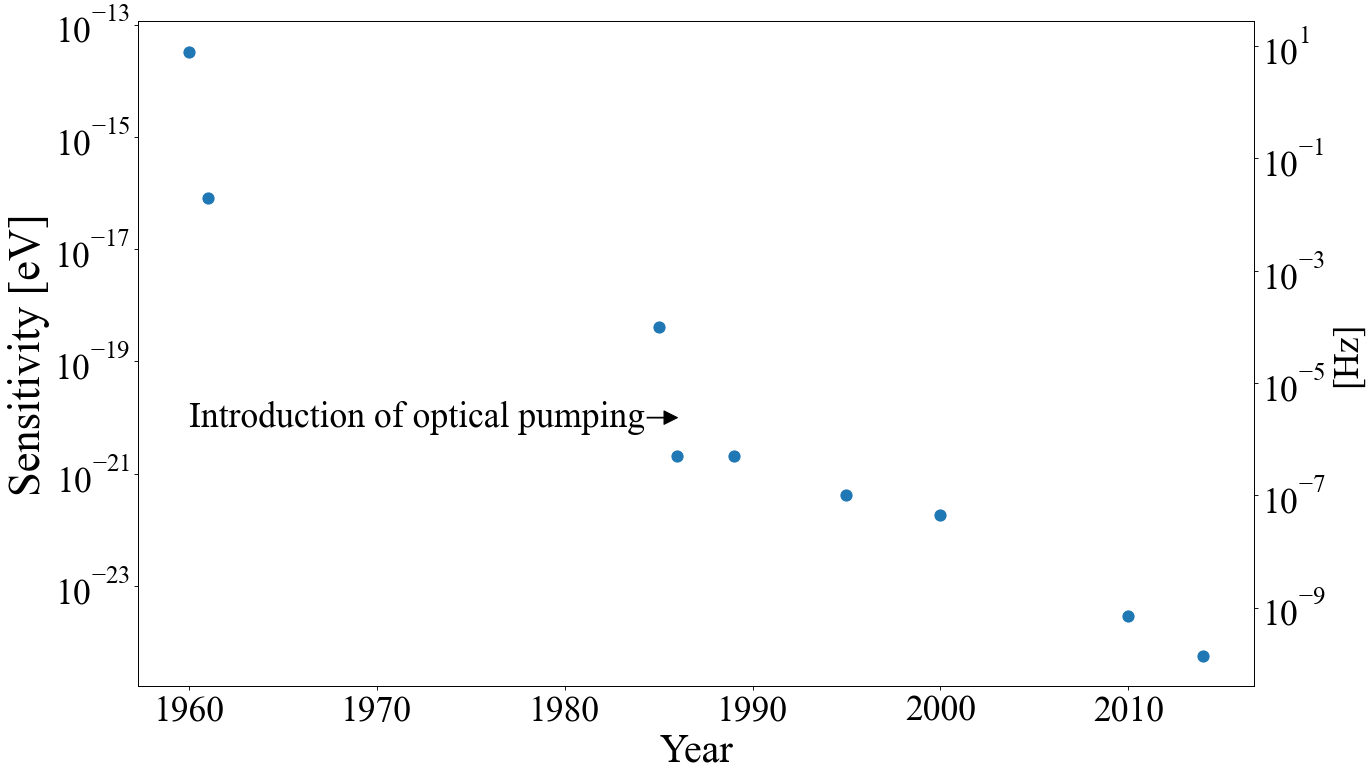}}
\caption{The progress in comagnetometer energy sensitivity since Hughes and Drever.  These results are for the energy of a spin due to its absolute orientation. The largest improvements came with the ability to create relatively pure ensemble quantum states via optical pumping, in the 1980s, and with the implementation of quantum magnetometers for the read-out systems in the 2000s.  References, from top-left to bottom-right: \cite{Hughes1960, Drever1961, Prestage1985, Lamoreaux1986, Chupp1989, Berglund1995, Bear2000, Brown2010, Allmendinger2014}
}
\label{fig: comag progress} 

\end{figure}

Since the era of Hughes and Drever, comagnetometry has improved in absolute energy sensitivity by 12 orders of magnitude. 
Figure \ref{fig: comag progress} shows the progress in measuring the energy of a nuclear spin pointing due its orientation in absolute space.   Comagnetometers built for other purposes\cite{Vasilakis2009, Graner2016} have achieved 
energy senstivities in the $\mathrm{10^{-26}\,eV}$ range.

\section{Comagnetometry}

A variety of comagnetometer 
implementations have been developed, with 
applications to angular-rotation sensing and fundamental physics.  This section provides a brief description 
of comagnetometers used in probes of fundamental physics.  This article -- and only to limit the 
scope -- focuses on 
comagnetometers that use at least one nuclear spin.  This means important electron-spin 
comagnetometers such as 
those used in electron EDM experiments, which compare the energies of various electronic spin 
states\cite{Baron2014, Cairncross2017, ACME2018, Chupp2019}, and spin pendulums which 
compare the interactions of electronic spin and electronic orbital angular momentum\cite{Heckel2008, 
Terrano2015} are left out.

State-of-the-art comagnetometers all use optical pumping -- in some fashion -- to generate highly nonthermal spin ensembles.  Schematically, optical angular momentum is transferred from a laser 
to an electronic state by optical pumping, and then on to the nucleus.  Depending on the transition 
energies of the atomic states and the wavelengths of available lasers,  some nuclear hyper-fine states can 
be pumped directly while 
others are polarized via spin-exchange collisions with more easily pumped atoms\cite{Dehmelt1957, Bloom1962, Walker1997}.   
 
\subsection{General comagnetometer considerations}

The main types of nuclear-spin comagnetometers are distinguished by comparison type (energy splitting 
or quantization axis), spatial distribution (overlapping or not), nuclear spin species of interest, and 
readout system.

\subsubsection{Comparison type: Clock vs quantization axis} \hfill\\
\emph{Clock comparisons}: The precession frequency $f_i$ of each spin ensemble $i$ is measured, giving 
its spin-up/spin-down energy splitting $\Delta E_\mathrm{spin} = h f_i$.  The $f_i$ are then combined in 
ratios or linear combinations which isolate 
$\mathcal{H}_\mathrm{BSM}$ as well as possible from $\mathcal{H}_\mathrm{mag}$ and other non-BSM 
contributions to $\mathcal{H}_\mathrm{spin}$.  Clock 
comparisons are mostly limited by state-dependent self-interactions of the spins and by back-action of the 
read-out system 
on the nuclei, necessitating precise and repeatable initialization of the ensemble state and quantum 
decoupling techniques. Specific clock-comparison based comagnetometers are discussed in sections 
\ref{sec: hg} and \ref{sec: hexe} and the references therein.   

\emph{Quantization-axis comparisons}: The quantization axes of two spin ensembles are compared to 
determine if there is a 
component of $\mathcal{H}_\mathrm{BSM}$ acting on one of the ensembles.  These systems are largely 
limited by nanoradian instability in the  
read-out and polarization systems, which also affect the quantization axes, and by trade-offs between 
signal size and polarization lifetime.  Specific systems based on quantization-axis comparison are discussed in section 
\ref{sec: alkali-noble}, and the references therein.   

\subsubsection{Spatial distribution: overlapped vs separated} \hfill\\
\emph{Spatially-overlapped spins}: The spin ensembles are contained in a single chamber, so that they 
nominally sample  the 
same volume and the same $\vec{B}$.  Since the ensembles are in the same chamber they must be  
different species.  
Surface effects, 
gradients in temperature and gradients in polarization decay mean the volume coverage can never be 
perfectly identical.  Sections \ref{sec: hg}, \ref{sec: hexe}, \ref{sec: alkali-noble} and references therein discuss  
spatially-overlapped comagnetometers in more detail.

\emph{Spatially-separated spins}: The spin ensembles are located in separate chambers.  This allows the 
comparison of spins 
of the same species, or spin species which require significantly different environmental conditions.  Since they are not in the same 
volume they do 
not sample the same magnetic field.  However, by using several cells drifts in both $\vec{B}$ and gradients of $
\vec{B}$ can be canceled. If the same 
species is used in all chambers, $\vec{\beta}$ must be different in the different chambers.  Section \ref{sec: hg} and references therein discuss spatially-separated comagnetometers in more detail.  

\subsubsection{Choice of Nucleus} \hfill\\
The nucleus must contain an unpaired nucleon, in order to have net spin.  All comagnetometers thus far have used 
isotopes with spin 1/2 or 3/2.  The spin 3/2 
nuclei have mass and 
charge quadrupoles allowing them to probe categories of interactions that are inaccessible to spin-1/2 
nuclei, although the charge quadrupole couples to electric field gradients and shortens the spin ensemble coherence time.  All competitive 
comagnetometers in recent decades have used either mercury or one of the three noble gases (helium, 
neon and xenon) that 
have stable isotopes of spin $\leq 3/2$ in reasonable abundance. Table 
\ref{tab: isotopes} summarizes the most commonly used nuclei.

 Mercury
can be pumped and probed directly with UV light, has a high vapor pressure at convenient temperatures 
and 
has remarkably long nuclear-spin lifetimes\footnote{In the hundreds of seconds, corresponding to many 
collisions} and {high-sensitivity} to {CP-violation} due to its nuclear structure\cite{Engel2013}.  Noble gases 
can have much higher densities than 
Hg vapors at room temperature. However at high densities they can only be hyper-polarized via 
fairly slow collisional-exchange-polarization techniques.  The spin-polarization lifetime must be much longer 
than the polarization time to achieve highly coherent ensembles.  This naturally leads experimental 
attention to the highly-inert noble gases; gases which form molecules depolarize too rapidly.

\begin{table}[htp]
\begin{center}
\captionsetup{width=.9\linewidth}
\begin{tabular}{lccc}
Isotope & Spin & Pumping & Phase \\
\hline
\iso{199}{Hg} & 1/2 & optical & vapor  \\
\iso{201}{Hg} & 3/2 & optical & vapor \\
\iso{3}{He} & 1/2 & SEOP/MEOP & gas \\ 
\iso{21}{Ne} & 3/2 & SEOP & gas \\ 
\iso{129}{Xe} & 1/2 & SEOP & gas \\ 
\iso{131}{Xe} & 3/2 & SEOP & gas
\end{tabular}
\caption[]{Commonly used nuclei in comagnetometry.  SEOP refers to spin-exchange-optical-pumping, wherein nuclear spin polarization is built up via collisions with an atom which can be optically pumped, typically potassium or rubidium. MEOP is metastability-exchange-optical-pumping, wherein a metastable state is optically pumped and then exchanged to the ground state during collisions.}
\label{tab: isotopes}
\end{center}
\vspace{-5mm}
\end{table}% 

\subsubsection{Read-out system}\hfill\\
 All read-out systems measure the magnetic moments of the nuclei.  For mercury, 
the transitions used to polarize the nuclei can also be used to read-out the nuclear-spin orientation by 
optical (Faraday) rotation of a linearly-polarized probe beam.   Noble gas systems have been built both with 
optical magnetometers and pick-up loop magnetometers. 

Optical magnetometers typically utilize the same alkali atoms as 
are used to polarize the nuclei, and have the virtue of nearly complete overlap between the magnetic 
sensor and the 
nuclear spins. In addition, the effective magnetic field experienced by alkali-metal spins is enhanced due to 
contact interactions with the nuclear spins.  This enhancement ranges from a factor of 5 for helium to $
\sim$500 for 
xenon\cite{Schaefer1989kappa, Walker1997}.  The tight coupling between the magnetometer and the nuclei 
leads to significant back-action by the magnetometer atoms on the nuclei, which is the main drawback of the approach 
(along with its sensitivity to 
optical alignment). 
Decoupling and stabilizing the read-out is also challenging, since any feedback or control pulses 
applied to the magnetometer also affect 
the co-located nuclei as well\cite{Limes2018a}.  

External magnetometers significantly reduce these disturbances to the spins -- both from back-action and 
from magnetometer feed-back and decoupling systems.  Optimally coupling an 
external magnetometer to the spins while minimizing magnetometer noise can be challenging 
technically, however\cite{Sachdeva2019}.

\subsection{Comagnetometer implementations for fundamental physics}

In this section we describe the general operating principles of comagnetometers used for 
fundamental physics.  
Currently there 
are three leading implementations with roughly comparable energy resolution: The Hg-EDM 
comagnetometer (Sec. \ref{sec: hg}), 
the alkali-noble 
gas self-compensating comagnetometer(Sec. \ref{sec: alkali-noble}), and the He-Xe-SQUID clock-comparison (Sec. \ref{sec: hexe}).  We also touch on several important  
implementations that paved the way to modern comagnetometry.  New 
concepts which are under development are mentioned in ``Future directions: Novel 
comagnetometers'' (Sec. \ref{sec: novel}).  Fundamental physics motivations, signatures and 
measurements are described 
in Section \ref{sec: measurements}.

\subsubsection{Mercury comagnetometers}\label{sec: hg}

The $ ^{199}\mathrm{Hg}$ - $^{201}\mathrm{Hg} $ comagnetometer, a spatially-overlapping clock 
comparison, was developed in 
1983\cite{Aleksandrov1983} and was the first comagnetometer to utilize optical pumping and optical 
readout.  The original apparatus 
was built to search for a dipole-dipole force between electrons and neutrons.  Improved versions were used 
to search for preferred reference frames\cite{Lamoreaux1986} and spin-gravity 
interactions\cite{Venema1992}.  

The  $ ^{199}\mathrm{Hg}$-$ ^{199}\mathrm{Hg}$ comagnetometer is unique in utilizing a single spin-
species, and as such uses spatially separated spin ensembles.  The design is 
completely intertwined 
with its application in searching for an EDM, and it is discussed in detail in section \ref{sec: Hg} on EDM searches.  Since its first implementation in 
1987 the sensitivity of this comagnetometer to an EDM has improved from {10$^{-26}\,e$-cm} to 
{$7\cdot10^{-30}\,e$-cm} \cite{Lamoreaux1987, 
Jacobs1993, Jacobs1995, Romalis2001, Griffith2009, Graner2016}.

Mercury/cesium comagnetometers are spatially-separated clock comparisons of $ ^{199}\mathrm{Hg}$ nuclei and  
Cs electrons, and were first built in 1995\cite{Berglund1995}. This avoids the quadrupolar $ 
^{201}\mathrm{Hg} $ nucleus and has been used to search for preferred frames\cite{Berglund1995, 
Peck2012} and for long-range 5th forces\cite{Youdin1996, Hunter2013, Hunter2014}.

\subsubsection{Noble-gas/noble-gas clock-comparison}\label{sec: hexe}
Clock comparisons between pairs of noble gases date back to the 1980s and the development of dual-nuclear-spin-species optical pumping via spin-exchange with rubidium\cite{Chupp1988, 
Chupp1989}, which 
underpins all implementations.  The first version compared ${}^3\mathrm{He}$ and ${}^{21}\mathrm{Ne}$ 
to study local Lorentz invariance\cite{Chupp1989}. It used a single chamber containing the two gases and 
alternated between a polarization period and 
a measurement period, with the measurement using NMR excitation and read-out techniques.  

The ${}^3\mathrm{He}$ - ${}^{129}\mathrm{Xe}$ maser, a spatially-overlapping clock-comparison, was built in 
the 1990s\cite{Chupp1994, Stoner1996, Bear1998} and used 
for used for EDM\cite{Rosenberry2001},  5th force\cite{Glenday2008} and 
preferred frame\cite{Bear2000} searches. The maser consisted of two chambers, one to generate a 
population inversion via optical pumping of the spins, and the other to provide readout and the positive 
feedback needed to maintain the masing.  The feedback was applied using pickup coils resonant to the two maser frequencies.  The atoms diffused between 
the two chambers via a small tube.

The ${}^{129}\mathrm{Xe}$ - ${}^{131}\mathrm{Xe}$ comagnetometer was built as a 
gyroscope\cite{Walker2011} and used to search for new 
forces\cite{Bulatowicz2013}.  It is a spatially-overlapped clock comparison with rubidium vapor in the same 
cell as the xenon.  Lasers pumped and probed the Rb directly, and the Xe via the Rb-Xe interaction.

The most recent all noble-gas comagnetometers were $\mathrm{{}^3\mathrm{He}}$ - ${}^{129}$Xe - SQUID 
systems\cite{Gemmel2010a, 
Gemmel2010b, Tullney2013, Allmendinger2014, Allmendinger2019, Sachdeva2019}.  The noble 
gases were polarized in a separate Rb spin-exchange optical pumping station and then were transferred to 
an evacuated 
measurement chamber.  The spins were monitored with a SQUID magnetometer that measured the total 
magnetic field produced by the 
gas cell. The precession frequencies of the two nuclei were separated and extracted in data analysis, and  
a $\vec{B}_0$ invariant frequency 
computed:

\begin{eqnarray}
\omega_\mathrm{inv} = \omega_\mathrm{Xe} - \omega_\mathrm{He}(\gamma_\mathrm{Xe}/\gamma_\mathrm{He} ),
\end{eqnarray}
where $\omega_{i}$ and $\gamma_{i}$ are the frequency and gyromagnetic ratios of species $i=\mathrm{He,
\,Xe}$.  

This technique is currently limited by the self-interactions of the nuclear spin-ensembles. Significant further 
improvement could 
be possible with improved quantum control and decoupling techniques, as discussed in section \ref{sec: future noble gas}.

\subsubsection{Alkali-metal/noble-gas self-compensating comagnetometer}\label{sec: alkali-noble}

These comagnetometers compare the spin-quantization axes of colocated spin ensembles, one spin being an alkali-metal vapor and the other a 
noble gas (specifically K-He and Rb-Ne).  They have 
been built in a variety of 
configurations and used to search for preferred reference frames and 5th forces.  The principles behind these comagnetometers are given in several 
publications\cite{Kornack2002, Kornack2005}.  The 
spins are polarized in-situ: a 
laser optically pumps the electronic spins of the alkali-metal, which in turn polarize the noble-gas nuclei via 
{spin-exchange} collisions.  When the external 
magnetic field
applied along the pump direction matches the magnetic field exerted by the nuclear spins on the alkali 
spins, the deflection of the quantization axis of 
the electrons relative to the pump beam is 
determined by the difference between the non-magnetic interactions of the 
electronic and nuclear spins. This cancelation of the external magnetic field is only effective at frequencies 
below the Larmor frequency of the nuclear spins. In 
that regime the alkali electronic spins experience such a small net 
magnetic field that broadening due to alkali-alkali spin-exchange collisions is eliminated, and the alkali-spin 
orientation can be 
measured very sensitively via the polarization rotation of a linearly polarized probe beam.  This read-out 
consists of measuring the projection of the alkali-spins along the probe beam axis, so mechanical changes 
in the relative alignment of the pump and probe 
beams can be an issue.  Designs aiming to ameliorate this mechanical sensitivity are under 
development.  Comagnetometers of this type have been used in searches for 5th forces\cite{Vasilakis2009, Lee2018, Almasi2020} and preferred 
frames\cite{Brown2010, Smiciklas2011}, and are well suited to dark matter direct 
detection\cite{Graham2018, Bloch2020}. 

\section{Fundamental Physics Results}\label{sec: measurements}

\subsection{Electric Dipole Moment Measurements}\label{sec: Hg}
Searches  for intrinsic electric dipole moments are among the most important precision tests of 
fundamental physics.  An intrinsic EDM -- which has yet to be observed in any system -- must violate T, and 
therefore CP, symmetry\cite{Purcell1950}. 
Searches for nuclear EDMs are motivated by two major outstanding questions in fundamental physics:  the 
strong-CP problem 
and the baryogenesis question.  The strong-CP problem arises from instanton anomalies that generically 
produce a CP-violating 
term in the presence of a non-zero quark mass\cite{THooft1976}.  This CP-violating term, whose 
coefficient ($\theta_\mathrm{QCD}$) is expected to be $\mathcal{O}(1)$, is measured to be less than 
10$^{-10}$. This discrepancy is the strong-CP problem.  
The baryogenesis question is: how did the universe come to contain more matter than antimatter?  
Generating a matter/{antimatter} asymmetry requires a process that simultaneously violates CP-symmetry, 
violates Baryon number conservation and  is 
out of equilibrium\cite{Sakhorov1967}.  No such process in the standard model is sufficiently strong, 
making the identification of additional {CP-violation} a key part of understanding how the Universe came to 
be. 

EDM measurements with nuclear-spin comagnetometers began in 1984 using $^{129}\mathrm{Xe}$
\cite{Vold1984} and shortly thereafter using $^{199}\mathrm{Hg}$\cite{Lamoreaux1987}, which has greater 
sensitivity to the strong-CP parameter $\theta_\mathrm{QCD}$ at equal experimental EDM sensitivity. 
Here we summarize the most recent measurements of the EDM of diamagnetic atoms; the field of EDM measurements is rich, and there are good recent 
reviews\cite{Yamanaka2017, 
Chupp2019}.

The $^{225}$Ra EDM is more sensitive to fundamental sources of CP violation than most atoms 
\cite{Engel2013, Chupp2014}. 
 The EDM of $^{225}$Ra was measured for the first time in 2015 using an 
optical dipole trap\cite{Parker2015} and in a follow-up measurement found to be less than 
$1.4\cdot10^{-23}\, e$-cm\cite{Bishof2016}.  
Co-magnetometry was not used in these first-generation experiments, but future experiments will use 
$^{171}$Yb atoms held in the same optical trap as a co-magnetometer

The most recent  $ ^{199}\mathrm{Hg}$ search used 4 chambers, each containing a vapor of $ 
^{199}\mathrm{Hg}$. Different 
electric fields were applied to the 4 chambers, 
allowing the experimenters to cancel fluctuations in the magnetic field and its linear and quadratic gradients 
while maintaining 
sensitivity to an EDM via appropriate linear combinations of the 4 measured frequencies. The $ 
^{199}\mathrm{Hg}$ nuclei were polarized by optical-pumping, and the precession rate of the 
nuclei in each chamber was 
measured by optical rotation of probe a laser.  The most recent measurement, the sixth published iteration, 
reached an 
EDM sensitivity of  $7\cdot10^{-30} \,e$-cm\cite{Graner2016}\footnote{The 
first paper\cite{Vold1984} set a limit of  $1\cdot10^{-26} e\mathrm{-cm}$ and pointed out the technique 
could potentially improve by 4 orders-of-magnitude.  The most recent measurements have nearly reached 
this target.}, setting the tightest constraints on $\theta_\mathrm{QCD}
$ and several other potential sources of CP-violation\cite{Graner2016, Chupp2014}.  
The measurement was limited by a combination of magnetic field gradients and redistribution of the liquid $
\mathrm{^{199}Hg}$ droplets.  

The most recent ${}^{129}\mathrm{Xe}$ EDM searches were noble-gas clock-comparisons (Sec. \ref{sec: 
hexe}) between  ${}
^{3}\mathrm{He}$ and ${}^{129}\mathrm{Xe}$ gases\cite{Allmendinger2019, Sachdeva2019}. The ${}
^{3}\mathrm{He}$ atoms should 
have a negligible EDM due to their small size\cite{Schiff1963} while the  $ ^{129}\mathrm{Xe}$  atoms 
could 
have a sizable EDM, depending on the high-energy origin of the CP-violation\cite{Flambaum2002, 
Chupp2014, Yamanaka2017, Chupp2019, Yanase2020, 
Fleig2021}. An electric field ($\vec{E}$) was applied to the cell 
and inverted every few minutes, so the nuclei experienced 
many $\vec{E}$ states per measurement.  The $\omega_\mathrm{inv}$ during each  $\vec{E}$ state was 
extracted, and then $
\omega_\mathrm{EDM}$ calculated by weighting by electric field and 
inverse-variance of $\omega_\mathrm{inv}$ and taking the mean.  This technique reached 
$1.4\cdot10^{-27} e$-cm and was limited by 
slow 
drifts in  $\omega_\mathrm{inv}$: the duration of each  $\vec{E}$-state was chosen to be shorter than the $
\omega_\mathrm{inv}$ drifts to eliminate systematics due to the drifts, but this limited the 
interrogation time of each electric field state.  The drifts seemed to originate from interactions between the 
nuclei, see Sec. \ref{sec: future noble gas} for more detailed discussion.

\subsection{Searches for Preferred Frames}
Hughes and Drever's seminal work\cite{Hughes1960, Drever1961} marked the first of many searches for preferred frames with 
spins, motivated by a variety of theoretical ideas including non-universal couplings of gravity and electricity\&magnetism,  Lorentz-violating scenarios
and the observation that CPT-violation generates preferred frames that could couple to spin\cite{Cocconi1958, Lightman1973, Coleman1997, 
Colladay1997, Kostelecky1999, Greenberg2002, ArkaniHamed2005}.  
Searches for preferred-frames using nuclei with charge-quadrupole-moments (spin-3/2 and higher) 
test the Lorentz-invariance of Maxwell's equations at much greater sensitivity\cite{Lightman1973, Haugan1979, 
Lamoreaux1986, Flambaum2017}
 than can be done with photons\cite{Mueller2003, Eisele2009}.   Similar experiments have been done looking for anisotropy of maximum attainable velocity for electrons \cite{Hohensee2013, Pruttivarasin2015, Sanner2019}), but they also do not reach sensitivities comparable to experiments with nuclei. 
Since the 
preferred frame is presumably fixed in the galaxy, the sensitive axis of the experiment must be modulated 
relative to the galaxy.  
Most experiments fix their sensitive axis in the lab and use the rotation of the Earth to modulate its direction 
relative to the galaxy. 
Other 
experiments use a rotation stage to change the orientation of the sensitive axis\cite{Brown2010, 
Smiciklas2011, Peck2012}. 

Searches for preferred frames with the alkali/noble gas comagnetometers (Sec. \ref{sec: alkali-noble}) used 
a rotary 
platform to rotate the entire experiment and move the BSM signal to a higher frequency\cite{Brown2010, 
Smiciklas2011}.  This type of comagnetometer has  
great initial sensitivity but has challenges with long-term stability due to its sensitivity to drifts in the optical 
alignment.  The K-He 
experiment 
reached a sensitivity of $3\cdot10^{-24}\ut{eV}$ (0.7 nHz) to a preferred spin-orientation, and the Rb-Ne 
experiment a sensitivity a few 
times larger for quadrupolar shifts.  These 
experiments were limited by imperfections in the inversion of the sensitive axis.  This becomes a systematic 
issue due to the gyroscopic effect of the rotation of the lab, 
which produces a frequency shift that depends on the angle of the sensitive axis relative to the Earth rotation 
axis.

A search for a preferred frame using the $^3\mathrm{He}$ - $^{129}\mathrm{Xe}$ - SQUID system (Sec. \ref{sec: hexe}) was made using the rotation of the 
Earth to modulate the orientation of the 
sensitive axis relative to the Galactic center.  This measurement was limited by self-interactions of the nuclei, which caused frequency drifts that the experimenters attempted to model and separate from 
the sidereal signature of new physics\cite{Gemmel2010b, Allmendinger2014}. 
There is disagreement both about the specific physical origin of these drifts, and how successful they were at disentangling the drifts from the signature 
of new 
physics\cite{Romalis2014, Allmendinger2014b, Terrano2019b}.

\subsection{5th Force}\label{sec: forces}
An ultra-low-mass, weakly interacting boson can generate a weak, macroscopic force.  Such particles are widely predicted and, 
if the particle is a pseudoscalar, would couple to the axial-current Lagrangian $
\mathcal{L}_{\mathrm{ax}}=g_{\mathrm{a} \bar{\psi} \psi} \partial_{\mu} a \bar{\psi} \gamma^{\mu} \gamma^{5} \psi
$ and mediate a new force 
coupled to spin.   Such particles -- variously referred to as axions or axion-like-particles -- are produced as the pseudo-Goldstone bosons of new, high-energy symmetries, just as pions are the pseudo-Goldstone bosons produced by chiral 
symmetry breaking.  The coupling of a pseudo-Goldstone boson to standard-model fermions is given by 
\begin{equation}\label{eq: symmetry}
g_{\mathrm{p}}=C_\mathrm{f}
m_{\mathrm{f}} / f_\mathrm{a} 
\end{equation}
 where $m_{\mathrm{f}}$ is the mass of the fermion, $f_\mathrm{a}$ is the energy scale of the symmetry breaking and $C_\mathrm{f}$ is a 
dimensionless coupling constant expected to be of order one. 
If the broken symmetry is not exact, like in chiral SU(2), the boson picks up a small mass { $m_{\mathrm{a}}=\Lambda^{2} / f_\mathrm{a} $ }
where $\Lambda$ is the explicit symmetry breaking scale. The interaction mediated by the boson is suppressed at distances 
larger than the Yukawa length of the boson $\lambda=\hbar /\left(m_{\mathrm{a}} c\right)$.  Searches for new, long-range spin-coupled forces are therefore a general way to search for new hidden symmetries\cite{Moody1984}. 

A pure pseudoscalar mediates a dipole-dipole (spin-spin) interaction:
 
\begin{equation}\label{eq: dipole}
\begin{aligned} V_{\mathrm{dd}} &=\frac{g_{\mathrm{p}}^{2} \hbar^{2}}{16 \pi m_1 m_2 c^{2} r^{3}}\left[\left(\hat{\boldsymbol{\sigma}}_{1} \cdot 
\hat{\boldsymbol{\sigma}}_{2}\right)\left(1+\frac{r}{\lambda}\right)-3\left(\hat{\boldsymbol{\sigma}}_{\mathbf{1}} \cdot \hat{\boldsymbol{r}}\right)
\left(\hat{\boldsymbol{\sigma}}_{2} \cdot \hat{\boldsymbol{r}}\right)\left(1+\frac{r}{\lambda}+\frac{r^{2}}{3 \lambda^{2}}\right)\right] e^{-r / \lambda}. 
\end{aligned}
\end{equation}

Here, $\hat{\sigma}_{1,2}$ are the spins of the two particles, $m_{1,2}$ their masses, and $\vec{r}$ the position vector between 
them. If the boson has a scalar coupling $g_\mathrm{s}$ in addition, it also mediates a CP-violating $\hat{\sigma}\cdot \hat{r}$ interaction
(sometimes called a spin-mass interaction):

\begin{equation}\label{eq: monopole-dipole}
V_{\mathrm{md}}=\frac{\hbar g_{\mathrm{s}} g_{\mathrm{p}}}{8 \pi m_1 c}\left[(\hat{\boldsymbol{\sigma}}_{1} \cdot \hat{\boldsymbol{r}})\left(\frac{1}{r 
\lambda}+\frac{1}{r^{2}}\right)\right] e^{-r / \lambda}.
\end{equation}			

In 5th force searches, the comagnetometer serves as the detector and a source of the new force is 
placed in the vicinity to modify the energies of the comagnetometer spins.  In searches for the dipole-dipole interaction (Eq. \ref{eq: dipole}) the 
source contains polarized spins.  In searches for 
the monopole-dipole interaction (Eq. \ref{eq: monopole-dipole}) the source is unpolarized matter.  The energy shift from the new 
force is typically modulated by varying the source distance or polarization.

Spin-mass interactions of nucleons have been studied since the 1960's, when a proton gravitational dipole 
moment was briefly 
claimed, before being ruled out\cite{Arlen1969}.  In comagnetometer searches for a new spin-mass force, 
a mass of large density is 
moved closer to and further from the comagnetometer, varying the magnitude of the interaction in Eq.
 \ref{eq: monopole-dipole} through its $\vec{r}$-dependence. 
Many spin-mass experiments using comagnetometers have been performed, optimized for various 
Yukawa ranges\cite{Wineland1991, 
Venema1992, Bulatowicz2013, Tullney2013, Lee2018}. 

Spin-spin interactions of neutrons were first studied using a $ ^{199}\mathrm{Hg} $-$ ^{201}\mathrm{Hg} $ 
comagnetometer and the polarized electrons in nearby magnetized material\cite{Aleksandrov1983}.  The 
neutron-neutron 
coupling was first studied using the $ ^{3}\mathrm{He}$ - $^{129}\mathrm{Xe} $ maser. A chamber filled with 
{high-polarization-density}  
$ ^{3}\mathrm{He}$  sourced the potential, and its polarization was inverted at regular 
intervals to flip the 
sign of the spin-spin interaction (Eq. \ref{eq: dipole})\cite{Glenday2008}. The best current limits come from a 
similar experiment 
performed with an  alkali-noble gas comagnetometer (K-He) as the detector\cite{Vasilakis2009}, which 
improved on the earlier
constraint by 3 orders of magnitude to set a limit on an anomalous nuclear spin-spin interaction at  $2\times10^{-8}$ of their magnetic interaction.

The exchange of spin-1 bosons can generate a greater variety of potentials than those given in Eqns. \ref{eq: dipole} and \ref{eq: monopole-dipole}, in 
particular long-range 
and velocity dependent interactions\cite{Dobrescu2006, Fadeev2019}.  Searches for these interactions can be 
optimized with different source geometries and motions.  For many Yukawa lengths, the best source is the spin-polarization of 
the Earth and its large rotational velocity\cite{Hunter2013}.  Using the Earth as a source requires specialized comagnetometer 
geometries, for which the mercury/cesium comagnetometer is being optimized.

\subsection{Dark Matter}
Axions or axion-like-particles are well-motivated extensions to the standard model of particles, as described in section \ref{sec: forces}.  In 
the following, ``axion'' refers to anything that couples to the axial-current, including all generic pseudo-Goldstone bosons, and 
``QCD axion'' refers to an axion that also couples to the QCD anomaly, and can thereby solve the strong-CP problem.  In 
addition to mediating new forces, axions would be produced in the early universe and make up some or all of the dark matter of 
the Universe.  Axionic dark matter consistent with cosmological observations can have a vast range of masses, roughly from $10^{-22}\,
\mathrm{eV}$ to  $10^{2}\,\mathrm{eV}$ in the simplest scenarios. Historically, dark matter axions which also solve the strong-CP problem have 
attracted the most attention, but many 
others have been proposed and there is no specific reason that the dark matter particles need to resolve other outstanding 
problems in particle physics.  

The relevant  standard model coupling is the pseudo-scalar Lagrangian discussed 
in section \ref{sec: forces}, but with the dark matter halo of the galaxy as the source. The non-relativistic Hamiltonian is then\cite{Vorobev1995, Graham2013} $$
H_{\mathrm{ax}} \sim g_\mathrm{p} a_{0} m_{\mathrm{a}}\left(\sum_j \vec{v}_j \cdot \vec{\sigma}_{\psi} \cos 
\omega_{j} t \right),
$$ where the summation is over modes $j$ of the axion field.  The nuclear-spin energy splitting is 
modulated by (i) the axion oscillation frequencies $\hbar \omega_{j}=E_{j}  = (m_\mathrm{a}c^2 + m_\mathrm{a}v_j^2/2)$, (ii) changes in the 
interference among 
the modes which occur on the coherence timescale set by the velocity dispersion of the axions $t_c \sim 2\pi/\omega_j(\Delta v/c)^{2} \sim 2\pi10^6/\omega_j$, and (iii) by experimentally 
controllable changes in the orientation of the fermion spin $\sigma_{\psi}$. The magnitude of the signal is 
proportional to 
$a_0 m_\mathrm{a}$, where $a_0$ is the amplitude of the axion wave at the time of the measurement.  On 
average, $a_0 m_\mathrm{a} = \sqrt{\rho_\mathrm{DM}(\hbar c)^3}$ where $\rho_\mathrm{DM}$ is the local 
dark matter density.  

Axions which  solve the strong-CP problem also induce oscillating EDM moments in nuclei\cite{Hong1990}; such axions couple directly to gluon fields and so have an unavoidable contribution their mass 
which puts them out of the frequency range of existing comagnetometers\cite{Blum2014}.  Going to higher frequencies requires 
resonant (NMR) techniques or axion-photon searches.   It is uncertain precisely which technique will be optimal for which mass ranges, but it seems 
plausible that the best strategies will be
comagnetometer based from $10^{-22}\ \mathrm{to\ 
10^{-13}\,eV}$ (10 nHz - 100Hz)\cite{Alonso2019}. 

Some dark-matter searches using comagnetometers have been performed\cite{Abel2017, Wu2020}, 
although so far with 
energy resolutions of a few $10^{-19}\,\mathrm{eV}$,  
significantly worse than that of state-of-the-art comagnetometers which reach $10^{-26}\,\mathrm{eV}$.  An axion search 
reaching an 
experimental sensitivity of $10^{-26}\,\mathrm{eV}$ would 
be able to probe axion symmetry scales of  $10^{11}\,\mathrm{GeV}$, assuming the coupling constant $C_\mathrm{f}$ defined in Section \ref{sec: 
forces} is one.   
This level is beyond even the most aggressive 
constraints inferred from stellar cooling, and within an order-of-magnitude or two of the axion symmetry 
scales currently probed by ADMX\cite{Braine2020}. 

\section{Future directions}

\subsection{Fundamental and practical limitations}
State-of-the-art comagnetometers are the most sensitive measurements of the energy difference between two quantum states, 
of any type, in terms of absolute energy sensitivity.  Different implementations are limited by different sources of instability, which are challenging and time consuming to identify and find solutions to -- especially when pushing against multiple 
sources of instability simultaneously.  However, these do seem to be practical rather than fundamental limitations, and there is room for significant further improvement in the fundamental sensitivity of these systems.  Of course, precision
measurements in the 
real world that are limited by fundamental noise are rare. 

\subsubsection{Sensitivity and limitations of the ${}^{199}\mathrm{Hg}$ comagnetometer}

The most recent  ${}^{199}\mathrm{Hg}$ measurement reached a total energy sensitivity of $22\,\mathrm{pHz \ 
(9\cdot10^{-26}\,eV)}$, and resolution of  $6.5\,\mathrm{nHz}$ from each 240$\,$s measurement\cite{Graner2016, Graner2017}.  This is a factor of 2-3 
larger than 
the signal-to-noise limit, as computed from the photon shot noise. The excess was attributed to a combination of magnetic-field 
gradients and migration of the  ${}^{199}\mathrm{Hg}$ droplets, which affected the distribution of the polarized nuclei.   
Making cells which can support high-coherence times is an art, and much effort has gone into increasing the coherence times and working life of 
the cells. It is unknown if, and by how much, they may be further improved.  The spin lifetimes in the best cells were 
600-1000 seconds, with a test cell with natural Hg reaching a coherence time of $1000$ seconds\cite{Graner2017}. 
If those coherence times could be 
matched in 4 cells, the integration time could be increased by a factor of $b = 3\sim5$ for an equivalent improvement in 
statistical sensitivity\footnote{Assuming the ratio of light and dark times remains constant.  So far laser related systematics have been below the statistical noise, but they are a source of concern.\cite{Graner2017}}.  Significant further improvement could 
be achieved by slightly increasing the temperature of the cells to obtain higher $^{199}$Hg density, provided 
that does not introduce new sources of leakage current, which can introduce magnetic fields correlated with reversal of the electric field in the cell\cite{Graner2016}.

\subsubsection{Sensitivity and limitations of noble-gas clock comparisons}\label{sec: future noble gas}

The intrinsic frequency resolution of a clock-comparison experiment is described by the Cramer-Rao lower 
bound (CRLB)\cite{Gemmel2010b} 
\begin{equation}
\sigma_\mathrm{f}^{2} \geq \frac{12}{(2 \pi)^{2}\left(A / \rho \right)^{2} T^{3}} C
\label{eq: CRLB}
\end{equation}
where $\sigma_\mathrm{f}$ is the uncertainty in [Hz], $A$ is the signal amplitude, $\rho$ is the (white-noise) amplitude spectral 
density in [$A/\sqrt{\mf{Hz}}$], $T$ is the observation time in [s] and ${C}$ is a dimensionless parametrization of the signal 
decay\footnote{C=1 for a constant signal, and C=1.7 for a signal that decays exponentially for one decay time.}. 

The signal amplitude $A$ (in magnetic field units) can be written as $A=c_f \mu_0 M_n$, where $M_n=\mu_n 
n P$ is the nuclear magnetization, given by the product of the nuclear magnetic moment $\mu_n$, nucleon  
density $n$ and nuclear spin polarization $P$, and $c_f$ is a dimentionless flux coupling factor that describes 
the magnetic field sensed by the pick-up coil. The frequency uncertainty $\sigma_f$ can be related to the 
effective magnetic field sensitivity $\sigma_B= 2 \pi \sigma_f/\gamma_n$. Combining these equations we 
obtain $$\sigma_B^2=\frac{2\rho^2}{T} \frac{6 C}{(c_f\gamma_n \mu_0 M_n T)^2}.$$ The first factor 
$(2\rho^2)/T$ gives the uncertainty in magnetic field measurement for a time $T$ using a magnetometer with field sensitivity $\rho$. The 
second factor on the order of $1/(\gamma_n \mu_0 M_n T)^2$ gives a dimensionless sensitivity gain factor 
due to nuclear spin precession that is proportional to the angle of nuclear spin precession in their own 
magnetization.  

A fundamental factor that can limit the sensitivity is spin-projection noise which, for spin-1/2 
systems, is given by\cite{Huelga1997}
$$(\sigma_f^\mathrm{SN})^2=\frac{C_\mathrm{SN}}{(2 \pi)^2 N T_2 T}$$  where $T_2$ is the decay time, $N$ the number of spins and 
$C_\mathrm{SN}$ a constant of order 5-10 that depends on the experimental protocol. Unlike alkali-metal magnetometers, nuclear spin 
magnetometers are usually not limited by spin-projection noise since they contain $10^{19}$ or more atoms.  In the scenarios considered here the projection-noise limit is below the Cramer-Rao lower bound in all cases\footnote{The speculative case is within a factor of a few of the projection-noise
limit.}.

Table \ref{tab: params} shows parameter values from the most recent $^3\mathrm{He}$- $^{129}\mathrm{Xe}$-SQUID 
measurement, along with some possible improvements.  Table \ref{tab: app params} gives the specific experimental configurations (geometries, polarizations and pressures) needed for each improvement. The measurement time $T_\mathrm{Meas}$ in the most recent experiment was chosen such that $T_\mathrm{Meas}\times \mathrm{Drift} 
\approx \sigma_\mathrm{f}$.  If the drifts were small 
enough that the measurement time could last the decay time, the Cramer-Rao bound would improve by an order of 
magnitude. This would require a three orders-of-magnitude reduction in the drifts.  The drifts seem to be 
caused by interactions among 
the nuclei\cite{Terrano2019b}; we provide more detail on them and how they may be controlled in the discussion around Eq. \ref{eq: selfint}.   The first two upgrades in Table \ref{tab: params} are 
realistic targets for near-term 
improvements. The last three are long-term targets which are technically challenging, although all 
technical requirements have been demonstrated in other experimental systems.  

In practice it may not be possible to reach best-ever 
levels on all parameters simultaneously as there can be trade-offs among them. In particular 
$T_\mathrm{Decay}$ is 
sensitive to the environment in many ways.   
Still, based only on signal-to-noise considerations, the Cramer-Rao bound can be significantly reduced. 

\begin{table}[htp]
\begin{minipage}{\textwidth}\begin{center}
 \vspace{5mm}
\captionsetup{width=.9\linewidth}
\begin{tabular}{lccccccc}
System &Cell Mag & Coupling & Readout  & Self Int. & T$_\mathrm{Decay}$ &  T$_\mathrm{Int} $ & CRLB ($\sigma_f$) \\
    &     [fT]        &  [geom.]  &[fT/$\sqrt{\mathrm{Hz}}$] &   [supp.]  &  [s]  &  [s]  &    [nHz] \\
  \hline
 HeXe-2019 & 1.3\E{6} & 4.4\E{-3} & 6 & 1.5\E{-3} & 8000 & 500  & 2.7\footnote{The experimentally measured uncertainty was 3$\,$nHz.}  \\
\hline
\multicolumn{8}{c}{Near-Term Targets: evolutionary progress} \\
\hline
Suppress Int & 1.3\E{6} & 4.4\E{-3} & 6 & 3\E{-6} & 8000 & 8000 & 0.17  \\
Geometry & 1.3\E{6} & 0.22 & 1 & 1\E{-7}  & 8000 & 8000 & 6\E{-3}  \\
\hline
\multicolumn{8}{c}{Long-Term Potential: modified protocols needed} \\
\hline
Best Noise &  1.3\E{6} & 0.22 & 0.1 & 1\E{-8} & 8000  & 8000 & 6\E{-4}  \\
Long Decay &  1.3\E{6} & 0.22 & 0.1 & 3\E{-9} & 18000  & 18000 & 1.7\E{-4}  \\
High Pol\footnote{Generating high polarizations of both species simultaneously would require separate pumping cells or advancements in dual-species hybrid pumping, as the optimal pumping conditions for high Xe density and high He density are quite different.} 
&  4\E{7} & 0.22 & 0.1 & 4\E{-12} & 18000  & 18000 & 6.3\E{-6}  \\
\end{tabular}
\caption[]{Experimental parameters of the 2019 Xe-EDM experiment\cite{Sachdeva2019} and possible
improvements.  Specific configurations needed to reach these levels are 
given in Table \ref{tab: app params} in the appendix.  ``Cell Mag'': The magnetic field at the surface of the 
cell for 
the limiting species. ``Coupling'': The ratio of field at the cell to field through the pick-up loop, which  
depends on the geometry of the system. ``Readout'': The noise in the read-out sensor.  This is often  
dominated by the magnetic noise of the local environment.  ``Self Int.'': The factor by which the self-interaction Hamiltonian (Eq. \ref{eq: selfint}) must be suppressed, by cancelling terms against each other, preparing the system with very small 
longitudinal component or decoupling that Hamiltonian. ``$T_\mathrm{Decay}$'': The decay time constant of the transverse amplitude\footnote{$A(t) = A_0 e^{-t/
T_\mathrm{Decay}}$}. ``$T_\mathrm{Int}$'' Measurement time for a single frequency extraction.  Taken 
to be $T_\mathrm{Decay}$ unless comagnetometer instabilities require shorter durations. ``CRLB'': The 
theoretical best resolution of the experiment, computed for a single $T_\mathrm{Decay}$ time.}
\label{tab: params}
\end{center}
\end{minipage}
\end{table}%

In the real-world, precision experiments are typically limited by environmental noise 
and systematics rather than fundamental limitations.  
It is not possible to predict whether all instabilities can be controlled well enough to reach the Cramer-Rao 
bound: we do not even know all the effects that may be important. We outline some likely sources of instability and what it might take to control 
them. 

\begin{itemize}
\item \emph{Longitudinal interactions}:  The drifts in current experiments are caused by interactions 
between the nuclei, described by spin-state dependent Hamiltonians 

\begin{equation}\label{eq: selfint}
\mathcal{H}_\mathrm{i} = \sum_{j} \mu_0 \alpha_{ij}\vec{\mu}_{i}\cdot\vec{\mu}_{j}
\end{equation}
where the $\vec{\mu}_{i, j}$ are the magnetic moments of nuclear spin species $i, j$, and the components of 
$\alpha_{ij}$ are coupling strengths on the order of $10^{-2}$.    
All components of $\alpha_{ij}$  are generically non-zero and not canceled in the comagnetometer 
frequency. The $i=j$ terms depend on the cell geometry.

This Hamiltonian produces a spin-up/down energy splitting if there is a non-zero component of 
either nuclear spin along $\vec{B}$, and causes a 
frequency drift as this ``longitudinal" component of the polarization decays away\cite{Limes2019a, 
Terrano2019b}. 
The drift can be suppressed by precise state initialization to ensure there is no longitudinal polarization, 
decoupling sequences\cite{Waugh1968}, and 
specially chosen cell 
geometries\cite{Limes2019a, Terrano2019b}. The suppression 
required can scale faster than the improvement in signal-to-noise ratio in some cases\footnote{Consider  improving some aspect of the experiment by a multiplicative factor 
of $b$.  How much would longitudinal interactions 
  need to be suppressed in order to take full advantage?  If the improvement is in the noise (or in signal, 
  if achieved through better spin-SQUID coupling), the longitudinal interactions must be suppressed by $1/b$ to maintain  
  the same integration time.   If the increase is in the number of spins -- through higher pressures or polarization fraction or both 
 -- the longitudinal interactions must be suppressed by $1/b^2$, as the higher polarization increases both the output signal and 
 the internal magnetizations that source the longitudinal self-interactions.  If the increase is in $T_\mathrm{Decay}$, the 
 longitudinal interactions must be suppressed by $1/b^{3/2}$ to take full advantage of the longer potential integration time.  If $T_\mathrm{Int}$ 
 increases relative to $T_\mathrm{Decay}$ the required suppression scales as  $1/b^{5/2}$ (in the $T_\mathrm{Int} \ll T_\mathrm{Decay}$ limit) 
 since a larger fraction of the longitudinal magnetization decays during the integration time}. The 
improvements outlined in ``Near-Term Targets'' of Table \ref{tab: params} require suppressing self-interactions 
by $\mathcal{O}(10^4)$ compared to current experiments.  Achieving the ``Long-Term Potential'' 
would require significantly greater control over the self-interacting portion of the Hamlitonian, especially if the 
cell magnetizations are increased.

\item \emph{Earth rotation effects}:  The nuclei precess in the non-inertial frame of the rotating Earth so they 
pick-up an apparent frequency shift of  $f_{\bigoplus} \sim 10\,\mu \mathrm{Hz}$ for a 
horizontal magnetic field at mid-latitudes\cite{Gemmel2010a}.  This shift depends on the angle between the 
magnetic field and the Earth's rotation axis, so anything that changes that angle also changes $
\omega_\mathrm{int}$.  Some specific potential issues include: tilts or twists of the apparatus from loading of the Earth's surface in the 
vicinity; changes in the background magnetic field; and changes in the orientation of the Earth's rotational 
axis. Secular changes in the Earth's rotation period 
 correspond 0.1pHz changes in $\omega_\mathrm{inv}$ as the Earth's rotation period itself changes.  
 These slow changes in Earth's rotation can be monitored and removed easily by 
 stellar observations\cite{Schreiber2003, Schreiber2004}.  Changes in the orientation of the Earth's rotation 
 axis are known from very long baseline interferometry and observed in ring-laser gyroscopes. In a typical 
 experimental configuration (mid-latitude \& horizontal holding field) these changes would contribute daily 
 fluctuations of $\sim 0.5\,\mathrm{pHz}$\cite{Schreiber2013}.
 If the holding field is aligned with the Earth's rotation axis\cite{Venema1992} these are suppressed to 
 tolerable levels (Table \ref{app: polar}).  
 
 \item \emph{Gyroscopic coupling to lab motions}:  Slow tilts of the lab due to tidal, atmospheric and local 
 loading can produce a rotation around the sensitive axis of the system, which would be picked up by the 
 same gyroscopic coupling that produces the Earth rotation effect described above.  Commercial tilt-meters have a 
 sensitivity around 7 nrad/$\sqrt{\mathrm{Hz}}$\cite{JewellTiltmeter}, a few times the resolution needed to monitor and subtract this 
 coupling in the most sensitive scenarios.  The success of ring-laser gyroscopes which are sensitive to twists 
 demonstrates that --  when care is taken and the apparatus affixed to the bedrock -- laboratory twists 
 can be suppressed to around these levels\cite{Schreiber2013}.

\item \emph{Transverse self-interactions}:  Since the cell is not spherical, there will be through-space 
Ramsey-Bloch-Siegert 
shifts of the nuclei on each other\cite{Allmendinger2014} caused by the magnetic field of the rotating (transverse) component of the nuclear polarization.  The signal sizes in the HeXe-2019 system should 
cause a transverse-magnetization-dependent frequency shift of $\sim500\,\mathrm{pHz}$ and scale 
quadratically with magnetization\cite{Terrano2019b}.  Since 
the rotating amplitudes are measured during data-taking this effect should be possible to account for metrologically.

\end{itemize}

\subsubsection{Sensitivity and limitations of alkali-noble gas self-compensating comagnetometers}

Alkali-noble gas comagnetometers face three major sources of noise, and the limiting source depends on the frequency and the 
measurement of interest:  at low-frequency, fluctuations in the beam alignment; at high-frequency, the breakdown of the magnetic-field 
compensation; in the middle, probe beam noise.  Experiments which require mechanical motion of the 
comagnetometer or source masses are often limited by beam alignment across the middle frequencies as well. 

The most sensitive comagnetometer ever operated was a K-${}^3$He co-magnetometer used to search for spin-spin 5th forces, 
which reached an integrated sensitivity of $20\,\mathrm{pHz \ (7\times10^{-26}\,eV)}$, with a noise level of $21\,\mathrm{nHz/
\sqrt{Hz}}$ at the signal frequency of 0.3\,Hz.  This experiment was limited by laser intensity fluctuations at a factor of two above  
the shot noise limit, and within a factor of 5 of the best-ever alkali-metal magnetometer sensitivity\cite{Dang2010}.

Beam alignment and magnetic field noise both improve at higher frequencies, so a system with larger magnetization which can
operate at higher-frequencies could be more sensitive.  The effect of mechanical motion can be reduced with pulsed operation of the pump laser followed by measurements of the transient response of the coupled spin system. Unlike nuclear spin precession magnetometers, the self-compensating magnetometer measures the twist on the nuclei at static equilibrium and so does not have the gain factor due to measuring the precession frequency given by $\gamma \mu_{0} M_n T$.   Thus its sensitivity is limited by the best available sensitivity of the alkali-metal magnetometer.

\subsection{Novel comagnetometer implementations}\label{sec: novel}

New ideas in comagnetometry are constantly being investigated, aiming for reduced sensitivity to various sources of systematic error.  
Here we give a brief description of some recent efforts, which explore several interesting avenues. 

\begin{itemize}
\item \emph{${}^{3}\mathrm{He}$ -$ {}^{129}\mathrm{Xe}/{}^{21}\mathrm{Ne}$ with Rb readout}\cite{Sheng2014, Limes2018a, Limes2019a}:  This system suppresses the longitudinal 
interactions that limit other noble-gas comagnetometers by using the same volume for pumping and probing, permitting  
greater refinement and repeatability of the state initialization at the beginning of the probe time.  It also increases the signal due to the contact interaction between Rb and nuclear spins.  The read-out of the spin precession is done via the Rb in the cell, meaning the back-action of the Rb on the nuclei must be decoupled.  This is done using RF-pulses 
to rapidly invert the Rb orientation so it has no net longitudinal polarization. These pulses also invert the Rb relative to the 
magnetic field, suppressing spin-exchange relaxation even at the higher magnetic fields needed for noble gas precession measurements.  The 
decoupling pulses themselves produce a species dependent frequency shift that is much smaller than that of the Rb which they 
decouple but still much larger than the target sensitivity.  This requires further decoupling, or operating in the dark with no 
pulses.   Interactions between Rb and ${}^{21}\mathrm{Ne}$ are $\sim 15$ times smaller than between Rb and ${}
^{129}\mathrm{Xe}$, allowing longer spin-precession times and smaller Rb back-action, at the cost of dealing with ${}
^{21}\mathrm{Ne}$ quadrupolar effects.  Current data shows the ${}^{21}\mathrm{Ne}$ quadrupolar frequency splitting is smaller than the  ${}
^{21}\mathrm{Ne}$ linewidth, even with decay times of several thousand seconds. 

\item \emph{Transversely pumped ${}^{129}\mathrm{Xe}$ - ${}^{131}\mathrm{Xe}$}\cite{Korver2015, Thrasher2019, Sorensen2020}:  This system suppresses the longitudinal 
interactions that limit other noble-gas comagnetometers by pumping the nuclei transversely to the holding field.   To pump the Rb perpendicular to the holding field, the magnetic ``holding" field is made up of pulses, with each pulse flipping the Rb by $2\pi$.    
To generate a net Rb polarization along the xenon as the xenon rotate, either the 
laser polarization was reversed at the xenon
frequencies, or the magnetic holding field was modulated so the xenon precess slower when aligned with the beam and faster when anti-aligned.  This pump/probe geometry also allows feedback to 
reduce the build up of longitudinal polarization.

\item{Dual Xe isotope spin maser}\cite{Sato2018}: This approach uses a Rb magnetometer to apply positive feedback so the nuclear 
spins precess indefinitely with continuous optical pumping. Compared to earlier work on spin masers [28,13], 
this approach can be operated at a lower bias field because it does not rely on inductive detection of spin 
precession with a pick-up coil. Even though the spins precess indefinitely due to the positive feedback, the 
sensitivity of spin masers is still given by the Cramer-Rao bound with $T$ equal to the spin coherence time. 
At longer times the feedback system introduces a random frequency walk due to coupling of the detection 
noise.

\item \emph{Comolecular comagnetometer}\cite{Wu2018, Wu2020}:  This system compares the ${}^{13}\mathrm{C}$ and ${}^{1}\mathrm{H}$ of liquid state $\mathrm{{}^{13}CH_{3}
CN}$ molecules to improve the spatial overlap between the comagnetometer spin-ensembles.   
These
experiments used the classic NMR technique of polarizing the nuclei in a large magnetic field.   The lack of 
hyper-polarization means the energy resolution of the first iteration was around $\mathrm{100\,\mu Hz}$ with 
30 days of integration time.  There is 
hope that hyper-polarization may be applied to this system\cite{Wu2020}.

\item \emph{Comagnetometer networks}:  Searches for transient anomolous fields are motivated by 
astrophysical models 
predicting things such as domain walls, axion vortices or self-gravitating axion clusters.  More generally, 
they are a good way 
to look for the unexpected\cite{Pospelov2013}. Identifying transients with a single 
comagnetometer is essentially impossible, as it would be 
indistinguishable from an experimental 
glitch.  A network of comagnetometers looking for correlated glitches across the globe, however, has a 
chance of observing such a transient\cite{Afach2021}.

\end{itemize}

\subsection{Potential Physics Reach}

Real physics experiments are limited by systematics, so projections of physics reach are inherently speculative.  However, to give an idea of what could be done, Table \ref{tab: scenarios} shows rough sensitivities to the axion-decay-constant and ${}^{129}$Xe EDM if the experimental targets outlined in Table \ref{tab: params} are reached, and control of systematics keeps pace. The  ``Near-term'' scenario corresponds to line 3 of Table \ref{tab: params} and involves canceling drifts and optimizing the experimental geometry.  ``Optimistic'' scenario corresponds to line 5 of Table \ref{tab: params}, and requires matching the best readout noise yet achieved in such a system, as well as the longest published polarization lifetimes.  ``Speculative'' scenario matches the highest spin-polarizations ever achieved -- which have not yet been demonstrated in dual-species systems -- along with much greater control over internal interactions, which is required because of the higher magnetizations.  It may also be possible to improve the readout noise of the system even further, which could be a more profitable approach, although just as speculative. 

\begin{table}[htp]
\begin{minipage}{\textwidth}\begin{center}
 \vspace{5mm}
\captionsetup{width=.9\linewidth}
\begin{tabular}{p{0.2\textwidth}p{0.3\textwidth}p{0.2\textwidth}p{0.2\textwidth}}
Experimental \newline progress & Integrated energy \newline resolution [eV (Hz)] & Dark-matter-axion scale $F_\mathrm{a}/C_\mathrm{n}$ [GeV] & Xe EDM [e-cm]  \\
\hline
Near-Term & 1\E{-27} (2.5\E{-13)}  & 9.6\E{11} & 2.9\E{-30} \\
Optimistic & 4\E{-29} (9.4\E{-15}) & 2.6\E{13} & 1.1\E{-31} \\
Speculative & 1.4\E{-30} (3.4\E{-16})& 7.1\E{14} & 4\E{-33} \\
\end{tabular}
\caption[]{Physics reach of a ${}^3\mathrm{He}$ - ${}^{129}\mathrm{Xe}$ - $\mathrm{SQUID}$ system under conservative, optimistic and speculative scenarios, assuming 100 days of measurement.   ``Integrated Energy Resolution'': The uncertainty on $\hbar \omega_\mathrm{inv}$.  ``Dark-matter-axion scale'':  Defined in equation \ref{eq: symmetry}, $F_\mathrm{a}$ is the symmetry-breaking scale of the axion and $C_\mathrm{n}$ is a dimensionless coupling constant to nucleons, assumed to be order one.  These limits apply specifically to axions with oscillation frequency below the repetition rate of the experiment, around  10$^{-19}\,\mathrm{eV}$.  ``Xe EDM [e-cm]'': The EDM measurements assume 25-33 kV voltage difference across the cell (5 kV/cm electric fields) with 5 high voltage states per decay time.  Leakage current effects are largely canceled by the comagnetometer, with measurements constraining an effect of $\leq 1.77\, \mu\mathrm{Hz}/\mu\mathrm{A}$.  This implies conservative estimates of the maximum allowable leakage currents to be 3.9\E{4}, 110 and 40 fA respectively.  The most recent ${}^{199}$Hg measurements measured steady-state leakage currents of 40 fA.}
\label{tab: scenarios}
\end{center}
\end{minipage}
\end{table}

\newpage

The sensitivities outlined in Table \ref{tab: scenarios} are ambitious, and will require significant advances in  
the understanding of and control over instabilities in this type of comagnetometer. They are, however, 
consistent with existing signal-to-noise ratios, which are outstanding thanks to decades of work.  

Enticingly, based on these estimates a comagnetometer could come within nearly an order-of-magnitude of some 
heretofore almost unimaginable targets: axion dark matter produced by symmetry-breaking at the Grand 
Unification scale 
of 10$^{16}\,\mathrm{GeV}$, and the standard-model prediction for the \xe EDM of $5 \times
10^{-35}\,e$-cm\cite{Chupp2019}.  
 If all goes well on the 
comagnetometry side, and another order-of-magnitude beyond Table \ref{tab: scenarios} is to be achieved, 
the next technological break-throughs -- whether in low-noise magnetic shielding, two-stage readout or elsewhere -- 
may be driven by these fundamental physics motivations.

\emph{Acknowledgements:}  This work was made possible by Princeton University and the Simons Foundation.

\newpage

\appendix \label{app: physical}
\section{Physical parameters of potential upgrades}

\begin{table}[htp]
\begin{minipage}{\textwidth}
\begin{center}
\captionsetup{width=.9\linewidth}
\begin{tabular}{l
ccc
cc
cc
}
System & \multicolumn{3}{c}{Dimensions [cm]} & \multicolumn{2}{c}{Xenon} & \multicolumn{2}{c}{Helium} \\ % $T_\mathrm{Decay}$  & $T_\mathrm{Meas}$  \\
  & Cell & Pickup & Cell-SQUID &% cell and squid dimensions 
  P   & pol &  % Xenon columns
  P   & pol   \\  % Helium columns
%  0.01 & [$\frac{\mathrm{fT}}{\sqrt{\mathrm{Hz}}}$] & [pHz/s] & [s] & [s]  \\
  &  Diam.  & Diam.  & Z  & [bar]  & [] & [bar]  & [] \\
\hline
 HeXe-2019 & 2.0  &  0.24  & 2.9 &  0.14 &  0.12 &  0.65 & 0.004  \\
\hline
\multicolumn{8}{c}{Near-Term Targets: evolutionary progress} \\
\hline
Suppress Int & 2.0  &  0.24  & 2.9 &  0.14 &  0.12 &  0.65 & 0.004 \\
Geometry & 5.0  & 1.4    &     4.1   &  0.14  &  0.12  &  0.65  &  0.004  \\
\hline
\multicolumn{8}{c}{Long-Term Potential: modified protocols needed} \\
\hline
Best Noise &  6.7     &     5.0      &    4.7      &     0.14    &   0.12   &    0.65   &   0.004   \\
Long Decay &  6.7     &     5.0      &    4.7      &     0.14    &   0.12   &    0.65   &   0.004   \\
High Pol  &  6.7     &     5.0     &     4.7     &      1.0    &    0.5    &    1.1    &   0.06   \\
\end{tabular}
\caption[]{Experimental parameters of the 2019 Xe-EDM experiment\cite{Sachdeva2019} and possible future improvements.  All parameters have been demonstrated, albeit not necessarily in a 
comagnetometer system. ``Cell-SQUID'' is the distance from the center of the cell to the center of the SQUID pickup loop.
 ``P'' and ``pol'' are the pressure and polarization 
fraction of each noble gas species.  }
\label{tab: app params}
\end{center}
\end{minipage}
\end{table}%

\section{Requirements on stability and monitoring}

\begin{table}[htp]
\begin{minipage}{\textwidth}\begin{center}
\vspace{5mm}
\captionsetup{width=.9\linewidth}
\begin{tabular}
{lccccccc}
Progression & \multicolumn{3}{c}{Typical B0} & \multicolumn{3}{c}{Polar B0}  \\
 & Earth Axis &  B$\perp$ & B0 Tilt & Earth Axis &  B$\perp$ & B0 Tilt  \\
  & [fraction of] &  [fT] & [$\mu$rad] & [fraction of] &  [fT] & [$\mu$rad] \\
\hline
  \vspace{0.5mm}
Near-Term & 0.4 & 3\E{3}  & 1 & 220 & 1\E{7} & 5\E{3} \\
Optimistic & 1.5 \E{-2} & 78  & 2.6\E{-2} & 8 & 3\E{5} & 1.5\E{2}  \\
Speculative & 5.4\E{-4} & 2.7 & 1\E{-3} & 0.3 & 1\E{4} & 5 \\
\end{tabular}
\caption[]{Stability requirements on environmental parameters so that associated frequency shifts from changes in the gyroscopic pickup of Earth's rotation remain below the statistical sensitivity.  
``Typical B0'' are the requirements for a convenient mid-latitude, horizontally aligned magnetic field.  ``Polar 
B0'' are the requirements if the magnetic field is aligned with the Earth's rotation axis to within 100$\mu$rad.  
``Earth Axis'' is the changing orientation of the rotation axis of the earth.  This is unavoidable and must be 
subtracted from the measured frequencies based on other observations and geophysical models.  This column gives the maximum allowable  fractional error in the model of the polar motion at the experiment site. ``B$\perp$'' 
gives the requirement on how well known the transverse fields at the cell must be known, per measurement point and assuming a 2$\mu$T holding field.  ``B0 Tilt'' gives the requirement on how well tilts along the B0 axis must be known.}
\label{app: polar}
\end{center}
\end{minipage}
\end{table}

\newpage

\bibliographystyle{apsrev4-1_custom.bst}
\bibliography{../../MainBib11_19} 

\end{document}